\documentclass[aps,prb,groupedaddress,twocolumn]{revtex4}
\usepackage{graphicx}
\usepackage{multirow}

\begin{document}
\title {Disorder-induced superconductor to insulator transition and finite phase stiffness in two-dimensional phase-glass models.}

\author{Enzo Granato}

\address{Laborat\'orio Associado de Sensores e Materiais,
Instituto Nacional de Pesquisas Espaciais, 12227-010 S\~ao Jos\'e dos
Campos, SP, Brazil}

\begin{abstract}
We study numerically the superconductor to insulator transition in two-dimensional phase-glass (or chiral-glass) models with varying degree of disorder. These models describe the effects of gauge disorder in superconductors due to random negative Josephson-junction couplings, or $\pi$ junctions.  Two different models are considered, with binary and  Gaussian distribution of quenched disorder, having nonzero mean. Monte Carlo simulations in the path-integral representation are used to determine the phase diagram and critical exponents. In addition to the usual superconducting and insulating phases, a chiral-glass phase  occurs for sufficiently large disorder, with random local circulating currents of different chiralities. A transition from superconductor to insulator can take place via the intermediate chiral-glass phase. We find, however, that the chiral-glass state has a finite phase stiffness, being still a superconductor,  instead of the Bose metal, which has been suggested by mean-field theory.

\end{abstract}
%\pacs{74.81.Fa, 73.43.Nq, 74.40.Kb, 74.25.Uv}

\maketitle

\section{Introduction}
Random gauge models of disordered superconductors, such as the gauge-glass model, have been widely used to study the vortex glass transition of disordered type II superconductors  \cite{fisher89,huse90}. Gauge disorder in this case arises from  the combined effect of geometrical disorder and the  applied magnetic field, leading to random phase shifts  in the Josephson junctions  coupling local superconducting regions. However, phase shifts can also  arise from the presence of negative Josephson couplings or $\pi$ junctions \cite{bula77,spivak91,sigrist95}, even in the absence of an applied magnetic field, and they can lead to different phase transitions and changes in the transport and magnetic properties\cite{kusma92,kawa95,kawa97,eg04b}. The  {\it phase glass} considered  in this work is a random gauge model, which has been introduced  by Dalidovich and Phillips\cite{dalidovich2002phase,phillips2003absence,phillips03} in an attempt to explain a metallic phase intervening between the superconductor and insulator phases in the zero-temperature limit, observed experimentally in many disordered superconducting films \cite{yazdani1995,chervenak2000}, even without an external magnetic field \cite{jaeger1989,christiansen2002evidence}. It  incorporates the effects of quantum fluctuations due to the charging energy $E_c$ of local superconducting regions and disorder in the Josephson-junction coupling $E_J$ between them, allowing for negative $E_J$.  The metallic phase, called a Bose metal \cite{das1999}, would be a physical realization of the glassy state in such a model, for a sufficiently larger degree of disorder.  Alternatively, the phase-glass model  could also be regarded as a quantum version of  the  chiral glass  model \cite{kawa95,kawa97,eg98,leeyoung},  or XY spin glass \cite{villain1977},  with varying degree of disorder, studied in the context of spin glasses and high-$T_c$ superconductors containing $\pi$ junctions. The chiral order parameter arises from the directions of the local circulating currents (Josephson vortices) introduced by the frustration effects of negative junctions. A chiral-glass phase  occurs in the ground state of such models for sufficiently large disorder. Although such a glass phase is unstable against  thermal fluctuations  in two dimensions \cite{kawa91,eg98,kostakino99}, it remains stable to quantum fluctuations at zero temperature  \cite{eg2017random}, below a critical value of the ratio $E_c/E_J$. 

Phase-glass models should also be relevant for recent experiments on superconducting thin films, nanostructured with a periodic pattern of nanoholes  and doped with magnetic impurities \cite{ zhang2019quasiparticle}. A simple model for phase coherence in these systems
consists of a Josephson-junction array, with the nanoholes corresponding to the dual lattice \cite{eg16,eg16R}. Since the magnetic impurities can introduce $\pi$ junctions \cite{bula77}  distributed randomly, the transition to the insulating phase as a function of doping can be described  by a  chiral-glass model with varying degree of disorder.  

The phase-glass model with a  Gaussian distribution of disorder with nonzero mean has been studied in detail analytically, and the glass state found for larger disorder was shown to correspond to a Bose metal, within a mean-field theory approach \cite{dalidovich2002phase,phillips2003absence,phillips03}. This result and its extension to the magnetic field-tuned transition \cite{wu2006}, provide a compelling description of  experiments  showing metallic behavior in superconducting thin films \cite{jaeger1989,yazdani1995,chervenak2000,tsen2016}, in terms of a minimum model with glassy behavior \cite{phillips2016}. However,  although results from mean-field theory tend to agree with those for models with infinite range interactions or in high dimensions, they are usually not reliable for  low-dimensional systems with short-range interactions. Thus, for the phase-glass model in two dimensions, further investigation  by numerical simulation is required to verify  the phase diagram, determine the  critical behavior, and test the prediction of a Bose metal phase. 

\begin{figure}
\centering
\includegraphics[width=\columnwidth]{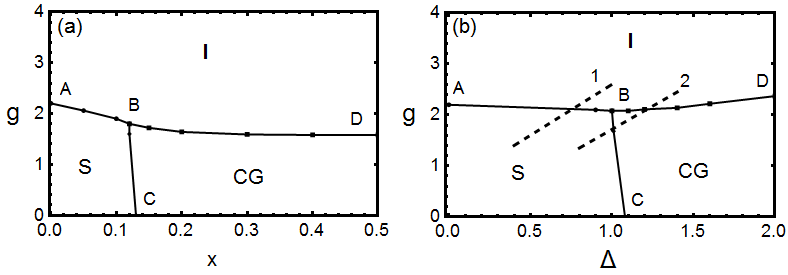}
\caption{Phase diagrams as a function of  $g=(E_c/E_J)^{1/2}$ and disorder strengths $x$ or $\Delta$, showing the superconducting (S), 
insulating (I) and chiral-glass (CG) phases. $E_c$ and $E_J$ are  the  charging 
and Josephson-coupling energies, respectively.  (a) Binary phase-glass model, where $x$ is the fraction of negative random Josephson couplings $\pm E_J$. (b) Gaussian phase-glass model, where $\Delta$ is the width of the Gaussian distribution of Josephson couplings with average value $E_J$.  The dotted lines $1$ and $2$ indicate two different possible paths for the superconductor to insulator transition.}
\label{phd}
\end{figure}

In this work, we study numerically the phase diagram and  critical behavior of two-dimensional phase-glass models with varying degree of disorder, at zero temperature. Two different models are considered, with a binary and  a Gaussian distribution of quenched disorder, both with nonzero mean. Monte Carlo simulations in the path-integral representation are used to determine the phase diagram and the critical behavior from
the finite-size scaling behavior of correlation lengths and phase stiffness. In addition to the usual superconducting and insulating phases, a chiral-glass phase (phase glass) occurs for sufficiently large disorder (Fig. \ref{phd}).  The transitions to the insulating phase across the lines AB and BD  are in different universality classes, with significantly different critical exponents and universal conductivity at the transitions. As sketched in Fig. \ref{phd}b, the transition from superconductor to insulator can take place as a single phase transition (path 1) or via the intermediate chiral-glass phase (path 2). We find, however, that the chiral-glass phase has a finite phase stiffness, being still a superconductor,  instead of the Bose metal, which has been predicted by mean-field theory \cite{dalidovich2002phase,phillips2003absence,phillips03} . This indicates that the simplest two-dimensional phase-glass model with short-range interactions does not provide a consistent theoretical framework for the Bose metal state in the zero temperature limit, as observed in recent experiments.
%These results indicate that the 2D phase-glass model  does not provided a theoretical framework for the Bose metal state, at least in its simplest form with short-range interactions and absence of magnetic screening effects. 

\section{Phase-glass models}
We consider disordered superconductors  described as a two-dimensional (2D) array of Josephson junctions, allowing for charging effects  and gauge disorder, defined by the Hamiltonian \cite{bradley84,fazio,phillips03,eg2017random}
\begin{equation}
{\cal H} = {{E_C}\over 2} \sum_i n_i^2 - \sum_{<ij>} E_{ij} \cos ( \theta_i -
\theta_{j}). \label{hamilt}
\end{equation}
The first term in Eq. (\ref{hamilt}) describes quantum
fluctuations induced by the charging energy, $ E_C  n_i ^2/ 2$,  of a
non-neutral superconducting "grain", or "island", located at site $i$ of a reference square lattice, where $E_C= 4 e^2/C$,
$e$ is the electronic charge, and
$n_i= -i \partial /\partial \theta_i $ is the operator, canonically conjugate to the phase operator $\theta_i$, representing the deviation of the number of Cooper pairs from
a constant integer value. The effective capacitance to the ground of each grain $C$
is assumed to be spatially uniform, for simplicity. The second term in (\ref{hamilt}) is the
Josephson-junction coupling between nearest-neighbor grains
described by phase variables  $\theta_i$. 
For a  spatially uniform Josephson coupling,  $E_{ij}=E_J > 0$. Equation  (\ref{hamilt}) is also known as the quantum rotor model \cite{cha3,sachdev2000}, with  additional effects of disorder in $E_{ij}$.

The  {\it phase-glass model},  as introduced  by Dalidovich and Phillips\cite{dalidovich2002phase,phillips2003absence,phillips03} to explain the Bose metal phase of superconducting films, incorporates the effects of disorder of $E_{ij}$ in Eq. (\ref{hamilt}) due to the random location of negative Josephson coupling ($E_{ij}<0$). Assuming an asymmetric  Gaussian distribution of $E_{ij}$ with nonzero zero mean,  this model has been studied in detail analytically, within a mean field theory approach.  

The phase-glass model can also be  regarded as a quantum version of  the 2D chiral glass  model \cite{kawa91,kawa97,eg98},  or XY spin glass \cite{villain1977},  with varying degree of disorder, studied in the context of spin glasses and high-$T_c$ superconductors containing $\pi$ junctions. In the classical limit $E_c \rightarrow 0$, the chiral order parameter arises from the directions of the local circulating currents (Josephson vortices) introduced by the frustration effects of negative junctions.  The chiral variable can be  defined as
\begin{equation}
\chi_{p} = \frac{1}{\chi_0}\sum^\prime_{<ij>} E_{ij} \sin (\theta_{i} - \theta_{j}),
\label{chiral}
\end{equation}
where the summation $\sum^\prime_{<ij>}$  is a direct sum around the plaquette $p$ of the lattice and $\chi_0$ is a normalization factor.
For sufficiently large disorder, a chiral-glass phase  occurs with random local circulating currents of different chiralities $\chi$. Although such a glass phase is unstable against  thermal fluctuations in two dimensions \cite{kawa91,eg98,kostakino99}, it remains stable to quantum fluctuations at zero temperature \cite{eg2017random}, below some critical value of  $E_c$. 

Since $E_{ij}<0$ is equivalent to a positive Josephson coupling 
$|E_{ij}|$ with a phase shift $A_{ij} = \pi $ to the phase difference $ \theta_i -
\theta_{j}$ in Eq. (\ref{hamilt}), the phase-glass model  is a particular case of random gauge models \cite{eg2017random}, with a binary distribution of phase shifts $A_{ij}=0$ or $ \pi $, in contrast to the  gauge-glass model \cite{stroud08,tang08}, where $A_{ij}$ has a continuous distribution.

We consider two different phase-glass models,  given by  asymmetric probability distributions of $E_{ij}$:
\begin{enumerate}
\item  $ P(E_{ij}) = x \delta(E_{ij}+E_J) + (1-x) \delta(E_{ij}-E_J)$     (binary) 
\item  $P(E_{ij}) =e^{-(E_{ij}-E_J)^2/2 \Delta^2 }/\Delta \sqrt{2 \pi}$  (Gaussian)
\end{enumerate}
The above binary and Gaussian disorder distributions are parameterized by $x$  and $\Delta$, respectively, with an average value of the Josephson coupling $<E_{ij}> \ne 0 $, except in the limit $x=0.5$ for the binary model, where $x$ corresponds  to the fraction of negative Josephson junctions. Only the  Gaussian phase-glass model has been studied in detail analytically, using a mean-field theory approach \cite{dalidovich2002phase,phillips2003absence,phillips03}.

\section{Path-integral representation and Monte Carlo simulation}
\label{MC}
The quantum phase transition at zero temperature can be conveniently studied in the framework of the  imaginary-time path-integral formulation of the model \cite{sondhi,sachdev2000}. In this representation, the 2D quantum model of Eq. (\ref{hamilt}) maps into a (2+1)D classical statistical mechanics problem. The extra dimension corresponds to the
imaginary-time direction. Dividing the time axis $\tau$ into slices $\Delta \tau$, the ground state energy corresponds to the reduced free energy $F$ of the classical model per time slice. The reduced classical Hamiltonian can be written as \cite{bradley84,wallin94,sondhi}
\begin{eqnarray}
H= &&-\frac{1}{g} [ \sum_{\tau,i}
\cos(\theta_{\tau,i}-\theta_{\tau+1,i}) \cr &&
+\sum_{<ij>,\tau} e_{ij}\cos(\theta_{\tau,i}-\theta_{\tau,j} ],
\label{chamilt}
\end{eqnarray}
%where a re-scaling of the time slices has been performed in order to obtain space-time isotropic couplings and
where $ e_{ij}=E_{ij}/ E_J$ and $\tau$ labels the sites in the discrete time direction. 
The ratio $g =(E_C/E_J)^{1/2}$, which drives the quantum phase transition  for the
model of Eq. (\ref{hamilt}), corresponds to an effective "temperature" in the
(2+1)D classical model of Eq. (\ref{chamilt}). The coupling of the phases $\theta_{\tau,j}$ 
in the time direction results from a Villain approximation, used  to obtain the phase representation
of the first term in Eq. (\ref{hamilt}), but it should preserve  the universal aspects of the critical behavior \cite{sondhi}.
%This approximation, however,  should preserve  the universal aspects of the critical behavior \cite{sondhi}. 
%In general, a quantum phase transition shows intrinsic anisotropic scaling, with  different diverging correlation
%lengths $\xi$ and $\xi_\tau$ in the spatial and imaginary-time directions, respectively, related by the dynamic critical exponent $z$
%as $\xi_\tau \propto \xi^z$.
%The energy gap $\Delta$ of the insulating phase is related
%to the phase correlation length in the time direction $\xi_\tau$ by $\Delta=1/\xi_\tau$.
The classical Hamiltonian of Eq. (\ref{chamilt}) can be viewed as a 3D  layered 
XY model, where  frustration effects exist
only in the 2D layers. Randomness in $e_{ij}$ corresponds to disorder
completely correlated in the time direction.

Equilibrium Monte Carlo (MC) simulations  are carried out using the
3D classical Hamiltonian in Eq. (\ref{chamilt}) regarding $g$ as a "temperature"-like parameter for different 
values of the disorder strength $x$ or $\Delta$. The parallel tempering method \cite{nemoto} is used in the simulations
with periodic boundary conditions, as in previous works \cite{eg16,eg16R,eg2017random}.
Since the correlation lengths in the spatial and imaginary-time directions are related by dynamical scaling
as $\xi_\tau \propto \xi^z$,
the finite-size scaling analysis is performed for different linear sizes $L$ of the square lattice with the constraint
$ L_\tau =a L^z $, where $a$ is a constant aspect ratio. This choice simplifies the scaling analysis, otherwise
an additional scaling variable $L_\tau/L^z$ would be required to describe the scaling functions.
The value of $a$ is chosen to minimize the deviations of $ a L^z$ from integer numbers. 
We used typically $ 10^4$ MC passes for equilibration and for calculations of average quantities. 
Averages over disorder used $100$ to $1000$ samples for system sizes  ranging from $L=16$ to $L=30$. 
Equilibration was checked with the methods described in Refs. \onlinecite{nemoto,bhat88}.
%However, this requires one to know the value of the dynamic exponent $z$ in advance.
%However, in presence of disorder, to estimate the value of $z$  with the above {\it equilibrium} MC method is %computationally very demanding. It requires performing simulations and disorder averages for increasing system sizes %$L_\tau$ in the time direction to find the optimum value of
%$z$ that gives the best scaling behavior.
%To overcome this problem, we follow a two-step approach.
%Since the exact value of $z$ is not known, we follow a two-step approach.
%First, we obtain an estimate of $g_c$ and $z$ from  simulations performed with a {\it driven} MC dynamics method, described below, which has been used in the context of the 3D classical XY-spin glass model \cite{eg04}. Then, these initial estimates are improved by finding the best data collapse for the finite-size behavior of the phase stiffness in the time direction $\gamma_\tau$, obtained by the equilibrium MC method. 

The MC simulations described above  employing periodic boundary conditions, do not allow a direct determination of the phase stiffness of the system in the spatial direction, $\gamma_x$, for large disorder. The dominant effect of the gauge disorder introduces additional phase shifts, which lead to negative values of phase stiffness depending on the disorder configurations. To probe the phase stiffness in this regime, we have also employed  a driven MC method with fluctuating boundary conditions \cite{eg04}. For that, the layered classical model of Eq. (\ref{chamilt}) is viewed as a 3D superconductor. In the presence of an external driving perturbation $J_x$ ("current density") that couples to the phase difference $\theta_{\tau,i + \hat x}-\theta_{\tau,i}$ along the $\hat x$ direction, the classical Hamiltonian of Eq. (\ref{chamilt}) is modified to
\begin{eqnarray}
H_J= H -\sum_{i,\tau} \frac{J_x}{g} (\theta_{\tau,i+\hat x}-\theta_{\tau,i}).
\label{driven}
\end{eqnarray}
MC simulations are carried out using the Metropolis algorithm and the time dependence is obtained from the MC time $t_{mc}$.
When $J_x \ne 0$,  the system is out of equilibrium since the  total energy is unbounded.
The lower-energy minima occur at phase differences $\theta_{\tau,i+\hat x}-\theta_{\tau,i}$,
which increase with time $t_{mc}$, leading to a net phase slippage
rate proportional to $ V_x = <d(\theta_{\tau,i+\hat x}-\theta_{\tau,i})/dt_{mc} > $, corresponding to the average
"voltage" per unit length.  The measurable quantity of interest is the phase slippage response ("nonlinear resistivity") defined as $R_x = V_x/J_x$. 
Similarly, we define $R_\tau$ as the phase slippage response to the applied perturbation $J_\tau$ in the layered (imaginary-time) direction.
%The  linear "conductivity" $\sigma = 1/R_x|_{J_x \rightarrow 0}$, is a measure of the phase stiffness of the system $\gamma_x$ througth the relation $\sigma(w)=\gamma_x(w)/i w $, where $w$ is the frequency. 
One then expects that  $R_x$ should approach a nonzero value when $J_x \rightarrow 0$ above the phase-coherence transition while below the transition it should  approach zero if the  phase stiffness is finite.  From the nonlinear "current-voltage" scaling  near the transition, one can extract the critical coupling $g_c$, and the critical exponents \cite{weng97,eg04,girvin,eg16R,eg2017random}.
%In the absence of charging effects, $R_x$ remains zero below a critical value $J_x=J_c$, which provides an
%estimate of the critical current for  the model of Eq. (\ref{hamilt}), when $E_c=0$.
%The scaling theory describing this behavior has been developed in the context of the current-voltage characteristics of %vortex-glass models \cite{vortexg}. In the present case, it needs to be generalized to take into account anisotropic %scaling, where $z\ne 1$. The required generalization has been described in detail in Ref. \onlinecite{girvin}.

\section{Numerical results and discussion}
The phase diagrams in Fig. \ref{phd}, were obtained by locating the S-I and CG-I transitions from the behavior of the correlation length and phase stiffness as a  function of $g$ for various fixed values of disorder strength, $x$ or $\Delta$.  The S-CG  transition was studied from the behavior of the correlation length as a function of disorder $x$ or $\Delta$ at fixed values of $g$.  In the following subsections, we described the results for the critical behavior across the transition lines AB, BC and BD at particular values  of  $x$ or $\Delta$, obtained by scaling analysis of extensive MC simulations. These results are summarized in Table I. The errorbars for the critical exponents are estimated from deviations
of the results obtained from different quantities.

\subsection{Superconductor to insulator transition }
\label{SI}

To locate the phase-coherence transition for a small degree of disorder, we first consider the behavior 
of the finite-size correlation length $\xi$, which can be defined as \cite{balle00}
\begin{equation}
\xi(L,g) = \frac{1}{2\sin(k_0/2)}[S(0)/S(k_0) - 1]^{1/2}.
\label{cdef}
\end{equation}
Here $S(k)$ is  the Fourier transform of the correlation function $C(r)$ and $k_0$ is the smallest nonzero
wave vector. For $g > g_c $, this definition  corresponds to  a finite-difference approximation to
the infinite system correlation length
$\xi(g)^2= -\frac{1}{S(k)} \frac{\partial S(k) }{\partial k^2} |_{k=0}$,
taking into account the lattice periodicity.  The correlation function in the spatial direction
is obtained as 
\begin{equation}
C(r) =\frac{1}{L^2L_\tau} \sum_{\tau,j} <\psi_{\tau,j} \psi_{\tau,j+r}>,
\label{cfunc}
\end{equation}
where $\psi_{\tau,j}=\exp( i\theta_{\tau, j})$ is the order parameter.  Analogous expressions 
are used for the correlation function $C_\tau(r)$ and correlation length $\xi_\tau(L,g) $ in the time direction.
For a continuous transition, $\xi(L,g)$ should satisfy the scaling form 
\begin{equation}
\xi/L = F(L^{1/\nu} \delta g ),
\label{correl}
\end{equation}
where $F(x)$ is a scaling function, $\delta g = g-g_c$ and $\nu$ is the correlation-length critical exponent. This scaling form implies that data for the ratio $ \xi(L,g)/L $ as a function of $g$, for different system sizes $L$, should cross at the critical coupling $g_c$.  Moreover, a scaling plot in the form $\xi/L \times L^{1/\nu} \delta g$  sufficiently close to $g_c$  should collapse the data on to the same curve.

\begin{figure}
\centering
\includegraphics[width=0.9\columnwidth]{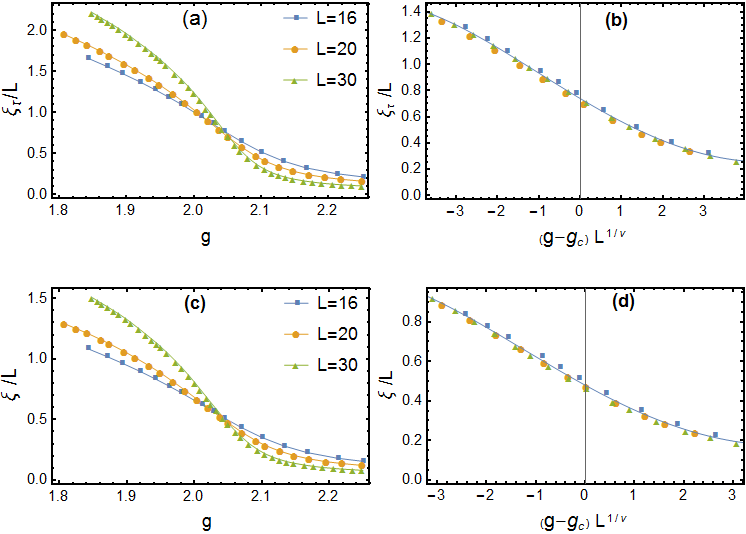}
\caption{(a) Correlation length in the imaginary time direction $\xi_\tau $ of the binary phase-glass model near the S-I transition at $x=0.05$ for  different system sizes $L$  and (b) corresponding scaling  plot of the data near the transition with  $g_c=2.048 $,  $\nu = 0.85 $.
(c) Correlation length in the spatial direction $\xi $ and (d) corresponding scaling  plot   with  $g_c= 2.05 $,  $\nu =0.89 $. 
$L_\tau=a L^z$, with aspect ratio $a=0.641$ and $z=1.1 $}
\label{corSI_pg}
\end{figure}

\begin{figure}
\centering
\includegraphics[width=0.9\columnwidth]{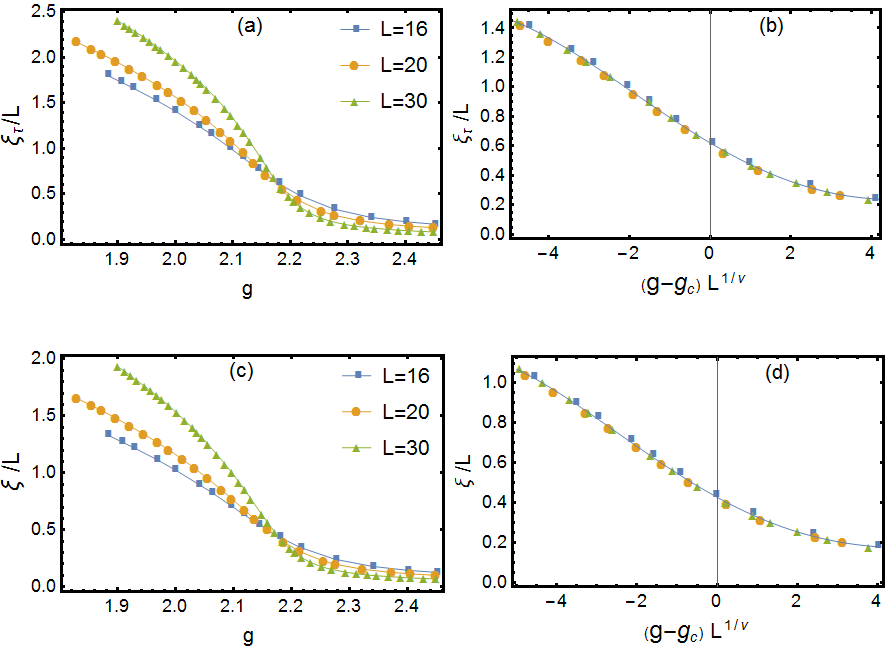}
\caption{(a) Same as Fig. \ref{corSI_pg}a but for the Gaussian phase-glass model at $\Delta=0.5$  and (b) corresponding scaling plot with  
$g_c=2.177 $,  $\nu = 0.86 $.
(c) Correlation length in the spatial direction $\xi $ and (d) corresponding scaling  plot   with  $g_c= 2.18 $,  $\nu =0.86 $. }
\label{corSI_gg}
\end{figure}

Figure  \ref{corSI_pg}a shows the behavior of the correlation lengths  $\xi_\tau $  and  $\xi$ in the time and spatial directions, for the binary phase-glass model with the dynamic exponent set to $z=1.1$.   The curves  for $\xi_\tau/L $ as a function of $g$ at a fixed value of disorder $x=0.05$ for different system sizes cross approximately at the same point, providing  evidence of a continuous transition. In  Fig. \ref{corSI_pg}b, a scaling plot according to Eq. (\ref{correl}) is shown, obtained by adjusting the parameters  $g_c$ and $\nu$ to obtain the best data collapse.  Figs. \ref{corSI_pg}c and \ref{corSI_pg}d show the corresponding behavior  for the correlation length in the spatial direction. The estimate $z=1.1$ was obtained by repeating the calculations for different values of $z$ larger than $z=1$ and choosing the one that gives the best data collapse.  Similar results are obtained for the  Gaussian phase-glass model as shown in Fig. \ref{corSI_gg}.

%Since in this case the crossing point is less clear, these estimates are more affected by corrections to finite-size scaling. For the flux and phase-glass models the difference of the estimate of $g_c$ from the correlation in the time and spatial directions are much larger.  We consider that the results obtained from the scaling of the phase stiffness $\gamma_\tau$ 
%% and correlation length $\xi_\tau$
%are more accurate and use them to obtain the final result and the associated errorbar. 

\begin{figure}
\centering
\includegraphics[width=\columnwidth]{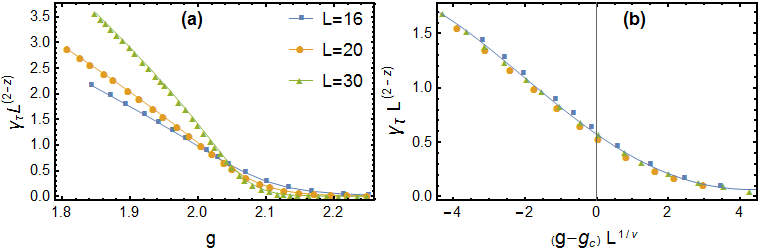}
\caption{ (a)  Phase stiffness in the imaginary time direction $\gamma_\tau $  of the binary phase-glass model near the S-I transition at $x=0.05$ for different system sizes $L$ and (b) corresponding scaling plot with  $g_c= 2.05 $  and $\nu = 0.82$.}
\label{helSI_pg}
\end{figure}

\begin{figure}
\centering
\includegraphics[width=\columnwidth]{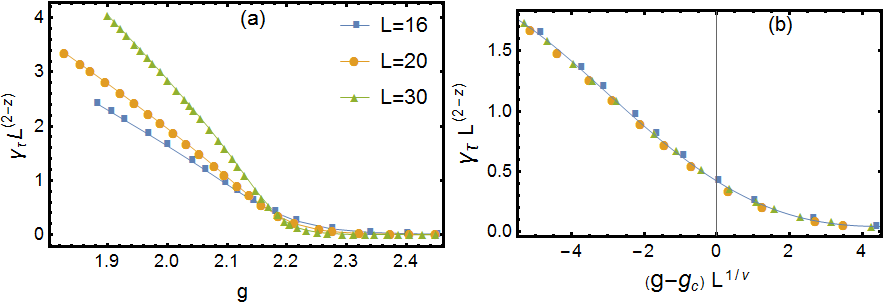}
\caption{ (a)  Same as Fig. \ref{helSI_pg}a but for the Gaussian phase-glass model at $\Delta=0.5$  and (b) corresponding scaling plot with  $g_c= 2.178 $  and $\nu = 0.84$.}
\label{helSI_gg}
\end{figure}

The phase-coherence transition described above can be identified as a superconductor-insulator transition from the behavior of the phase stiffness $\gamma$, which measures  the free energy cost to impose an infinitesimal phase twist along a certain direction. In  the imaginary time direction, $\gamma_\tau$, which  corresponds to the compressibility of the bosonic system,  is  given by \cite{cha2,wallin94}
\begin{eqnarray}
\gamma_\tau=\frac{1}{L^2 L_\tau g^2}[g <\epsilon_\tau>
- < I_\tau^2 > + < I_\tau >^2]_d,
\label{eqstiff}
\end{eqnarray}
where $\epsilon_\tau = \sum_{\tau,i} \cos(\theta_{\tau,i} -\theta_{\tau+1,i})$ and   $I_\tau= \sum_{\tau,i}\sin(\theta_{\tau,i} -\theta_{\tau+1,i})$. In Eq. (\ref{eqstiff}), $< \ldots>$ represents a MC average
for a fixed disorder configuration, and $[  \ldots ]_d$ represents an average over different disorder configurations. Similarly, the phase stiffness in the spatial direction,  $\gamma_x$, which corresponds to the superfluid density,  is given by the analogous expression in the $\hat x$-direction.  In the superconducting phase, $\gamma$ should be finite, reflecting the existence of phase coherence, while in the
insulating phase it should vanish in the thermodynamic limit.
For a continuous phase transition, $\gamma_\tau$ should satisfy the finite-size scaling form \cite{wallin94}
\begin{equation}
\gamma_\tau L^{2-z} = F(L^{1/\nu} \delta g),
\label{rhotau}
\end{equation}
where $F(x)$ is a scaling function and $\delta g = g-g_c$. This scaling form implies that data for $ \gamma_\tau L^{2-z} $ as a function of $g$, for different system sizes $L$, should cross at the critical coupling $g_c$. Fig. \ref{helSI_pg}a shows this crossing behavior and Fig. \ref{helSI_pg}b shows the corresponding  scaling plot of the data according to the scaling form of Eq. (\ref{rhotau}). The same  behavior  is found for the Gaussian phase-glass model as shown in Fig.  \ref{helSI_gg}.

\begin{figure}
\centering
\includegraphics[width=1.0\columnwidth]{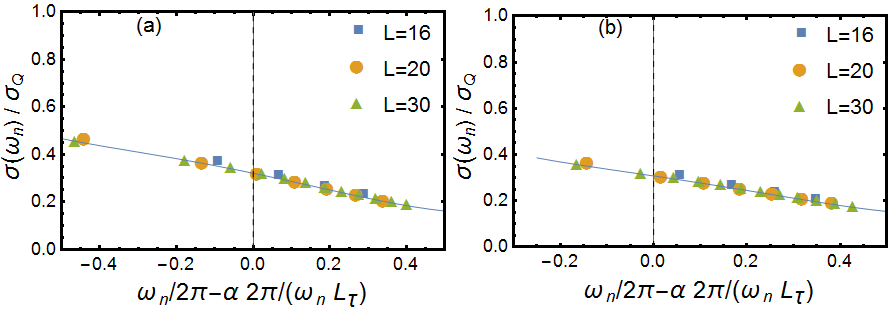}
\caption{  Scaling plot of conductivity $\sigma(iw_n)$ at the critical coupling $g_c$ for (a) the binary phase-glass model at $x=0.05$,
 $g_c=2.058 $ with $\alpha=0.5$ and (b) the Gaussian phase-glass model at $\Delta=0.5$, $g_c=2.187$ with $\alpha=0.2$. The universal conductivity  is given by the intercept with the $x_o=0$ dashed line, leading to the corresponding estimates (a) $\frac{\sigma^*}{\sigma_Q}= 0.32(3)$  and (b) $\frac{\sigma^*}{\sigma_Q}= 0.31(3)$. }
\label{cond-pg}
\end{figure}

Another quantity characterizing the superconductor to insulator transition is the electrical conductivity at the transition. Its value should be universal, it  does not depend on the parameters of the model, but it can depend on the  universality class of the transition \cite{fisher90a}.  Following the scaling method described by Cha {et al.} \cite{cha2,cha3}, the universal conductivity can be determined  from the frequency and finite-size dependence of the phase stiffness $\gamma(w)$ in the spatial direction. The conductivity is given by the Kubo formula
\begin{equation}
\sigma = 2 \pi \sigma_Q \lim_{w_n\rightarrow 0}  \frac{\gamma(i w_n)}{w_n},
\label{kubo}
\end{equation}
where $\sigma_Q=(2 e)^2/h$ is the quantum of conductance and $\gamma(i w_n)$ is a frequency dependent phase stiffness
evaluated at the finite frequency $w_n=2 \pi n /L_\tau$, with $n$ an integer. The frequency dependent  phase stiffness in
the $\hat x$ direction is given by
\begin{eqnarray}
\gamma(i\omega_n)=\frac{1}{L^2 L_\tau g^2} [& g &< \epsilon_x >
-< |I(i w_n)|^2 > \cr
& + & < |I(i w_n)| >^2]_d,
\label{helw}
\end{eqnarray}
where
\begin{eqnarray}
\epsilon_x&=&\sum_{\tau,j}  e_{j,j+\hat x}\cos(\Delta_x \theta_{\tau,j}) ,\cr
I(i w_n)&=&\sum_{\tau,j}  e_{j,j+\hat x} \sin(\Delta_x \theta_{\tau,j}) e^{i w_n \tau},
\end{eqnarray}
and $\Delta_x \theta_{\tau,j}= \theta_{\tau,j}-\theta_{\tau,j+\hat x}$.  
At the transition, $\gamma(i w_n)$ vanishes linearly with frequency and $\sigma$ assumes a universal value  $\sigma^*$, which can be  extracted from its frequency and finite-size dependence as  \cite{cha2}
\begin{equation}
\frac{\sigma(iw_n)}{\sigma_Q} = \frac{\sigma*}{\sigma_Q}
- c (\frac{w_n}{2 \pi} - \alpha \frac{2 \pi}{w_n L_\tau}) \cdots
\label{cond}
\end{equation}
The parameter $\alpha$ is determined from  the best data collapse of the frequency
dependent curves for  different systems sizes  in a plot of $\frac{\sigma(iw_n)}{\sigma_Q}$ versus
$x_o=(\frac{w_n}{2\pi} - \alpha \frac{2\pi}{w_n L_\tau})$. The universal conductivity is
obtained from the intercept of these curves with the line $x_o=0$. From this scaling behavior, shown in Fig. \ref{cond-pg}a  for the binary phase glass model and Fig. \ref{cond-pg}b for the Gaussian phase-glass model, we obtain at the S-I transition  $\sigma^*/\sigma_Q = 0.32(3)$ and $\sigma^*/\sigma_Q = 0.31(3)$, respectively.

\subsection{Superconductor to chiral-glass transition}

Above a disorder-strength threshold $x_B$, the finite-size correlation length $\xi(L,g)$ defined in Eq. (\ref{cdef}) no longer displays a crossing behavior at some critical $g_c$, indicating the absence of long-range  or quasi-long range order in terms of the order parameter  $\psi_{\tau,j}=\exp( i\theta_{\tau, j})$. From this change of behavior, we  estimate the location of the multicritical point B at $(g_B, x_B)$ for the binary phase-glass model in the phase diagram of Fig. \ref{phd}a  and, similarly, for the Gaussian phase-glass model at $(g_B,\Delta_B)$ in Fig.   \ref{phd}b.

\begin{figure}
\centering
\includegraphics[width=0.9\columnwidth]{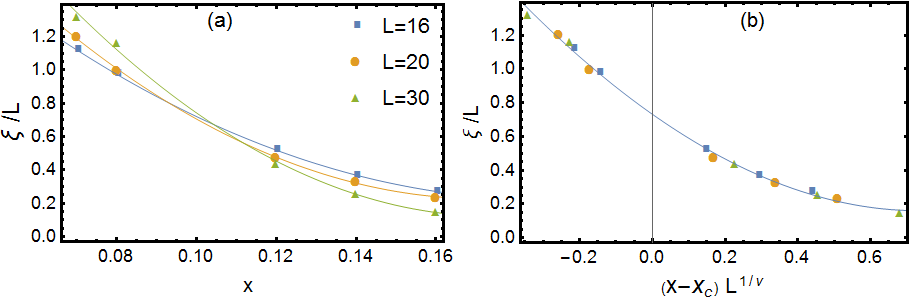}
\caption{(a) Correlation length in the spatial direction $\xi $ of the binary phase-glass model near the S-CG  transition at fixed $g=1.6$ for different system sizes $L$   and (b) corresponding scaling  plot of the data near the transition with  $x_c=0.1 $,  $\nu = 1.4 $.}
\label{corSCG_pg}
\end{figure}

\begin{figure}
\centering
\includegraphics[width=0.9\columnwidth]{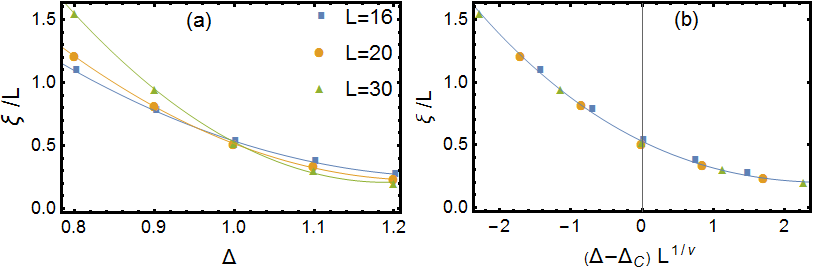}
\caption{(a) Same as  Fig. \ref{corSCG_pg}a but for the Gaussian phase-glass model near the S-CG  transition at fixed $g= 1.68$ for different system sizes $L$   and (b) corresponding scaling  plot of the data near the transition with  $\Delta_c=1.0 $,  $\nu = 1.4 $.}
\label{corSCG_gg}
\end{figure}

The superconductor to chiral glass transition can be determined from the behavior of the correlation length as a function of disorder at a fixed value of $g < g_B$.  Figure  \ref{corSCG_pg}a shows the behavior of the correlation length  $\xi(x,L)$  for increasing disorder $x$ at  $g = 1.6$, for the binary phase-glass model.   The curves  of $\xi(x,L)/L $ for different system sizes cross approximately at the same point $x_c$, providing  evidence of a continuous S-CG transition. In Fig.\ref{corSCG_pg}b, a scaling plot according to Eq. (\ref{correl}) is shown, obtained by adjusting the parameters  $x_c$ and $\nu$ to obtain the best data collapse. The dynamic exponent $z=1.1$ was obtained by repeating the calculations for different values larger than $z=1$ and choosing the one that gives the best data collapse.  Similar results are obtained for the  Gaussian phase-glass model as shown in Fig.  \ref{corSCG_gg}.

\begin{figure}
\centering
\includegraphics[width=\columnwidth]{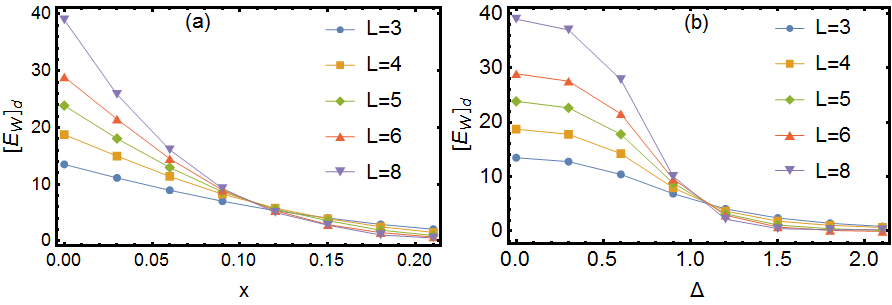}
\caption{Domain-wall energy $ [E_W]_d $ for the (a) binary and  (b) Gaussian phase-glass models in the limit $g  \rightarrow 0$ for increasing disorder ($x$ or $\Delta$) and different systems sizes $L$. The common crossing point of the curves gives an estimate of  the chiral-glass disorder threshold $x_C$ and $\Delta_C$ in the phase diagrams of Fig. \ref{phd}. }
\label{dw}
\end{figure}

Point C in the phase diagrams of Fig. \ref{phd} corresponds to the superconductor to chiral-glass threshold in the limit $g  \rightarrow 0$. It can be estimated from the finite-size behavior of the domain wall energy \cite{bray1984,benakli1998} in the ground-state of the 3D classical model of Eq. (\ref{chamilt}).  A domain wall in the finite system can be introduced  by imposing antiperiodic boundary conditions in one of the spatial directions.  The domain-wall energy $ E_W ( L) $  is a measure of
phase coherence, and is  related to the renormalized stiffness constant $\gamma_x(L)=   E_W(L)/(2\pi^2 L_\tau) $.  Although $ E_W ( L) $ fluctuates between samples with different disorder configurations, stability of the ground state with long-range order in  $\psi_{\tau,j}=\exp( i\theta_{\tau, j})$ requires that the disorder average $[E_W]_d$
increases with $L$ or remains constant. We have determined numerically the  change $ E_W ( L) $ in the ground-state energy of small systems
by MC simulated annealing  for a large number of samples. 
Figure \ref{dw} shows the behavior of the $[E_W]_d$ for different system sizes and  increasing disorder, for both  phase-glass models.  For small disorder strength, it increases with $L$, indicating the existence of long-range phase
coherence.  For sufficiently large disorder it clearly decreases for increasing $ L$, indicating a disordered glass phase. The change in
the behavior yields the  estimates of  the chiral-glass disorder thresholds $x_C= 0.11(2) $ and $\Delta_C=1.1(2)$, for the binary and Gaussian phase-glass models, respectively. 

\subsection{Chiral glass to insulator transition}
\label{CGI}
In the chiral-glass phase for $x > x_B$ or $\Delta >\Delta_B$ in Fig. \ref{phd}, there is no long-range order in terms of the order parameter $\psi_{\tau,j}=\exp( i\theta_{\tau, j})$. It is then convenient to define a glass correlation function in
terms of the overlap order parameter \cite{bhat88,rieger1994zero} of phase variables $q_{\tau,j}=\exp( i(\theta^1_{\tau, j} - \theta^2_{\tau,j}) )$, 
where $1$ and $2$ label two different copies of the system with the same coupling parameters. The glass correlation function in the spatial direction
is then obtained as 
\begin{equation}
C(r) =\frac{1}{L^2L_\tau} \sum_{\tau,j} <q_{\tau,j} q_{\tau,j+r}>,
\label{cfunc}
\end{equation}
with the  corresponding phase correlation length $\xi_{G}$ defined as in  Eq. (\ref{cdef}). Analogous expressions are used for the correlation length in the time direction $\xi_{G,\tau}$. Similarly, we can also define a chiral correlation length $\xi^c_G$ in terms of the overlap of  the chiral variables of
Eq. (\ref{chiral}), $q^c_{\tau,p}=\chi^1_{\tau, p} \ \chi^2_{\tau,p} $.

\begin{figure}
\centering
\includegraphics[width=0.9\columnwidth]{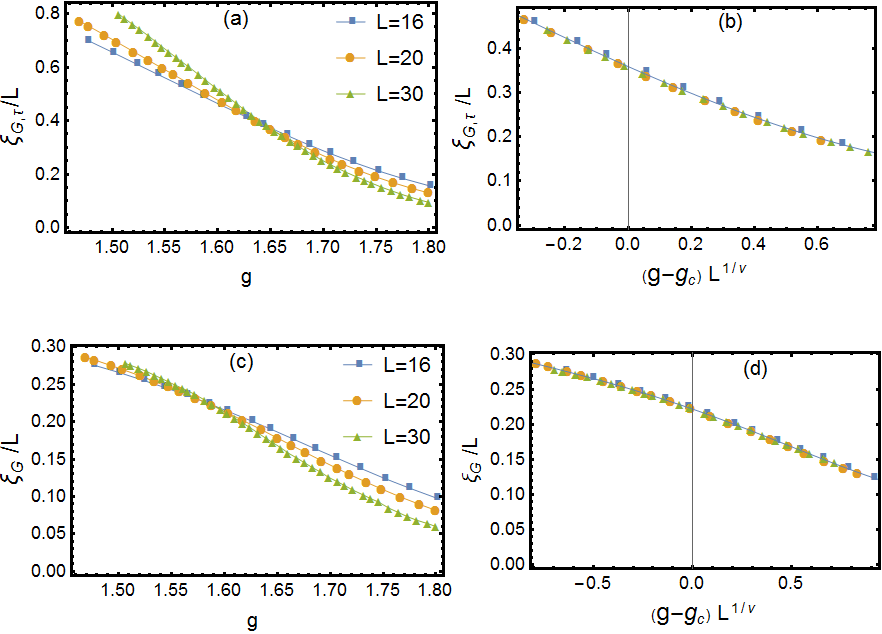}
\caption{(a) Correlation length in the imaginary time direction $\xi_{G,\tau} $  of the binary phase-glass model near the CG-I transition at $x=0.2$ for  different system sizes $L$   
and (b) corresponding scaling  plot of the data near the transition with  $g_c=1.655  $,  $\nu = 1.6 $.
(c) Correlation length in the spatial direction $\xi_G $ and (d) corresponding scaling  plot   with  $g_c= 1.59 $,  $\nu = 1.6 $. 
$L_\tau=a L^z$, with aspect ratio $a=0.642$ and $z=1.2$}
\label{corCGI_pg}
\end{figure}

\begin{figure}
\centering
\includegraphics[width=0.9\columnwidth]{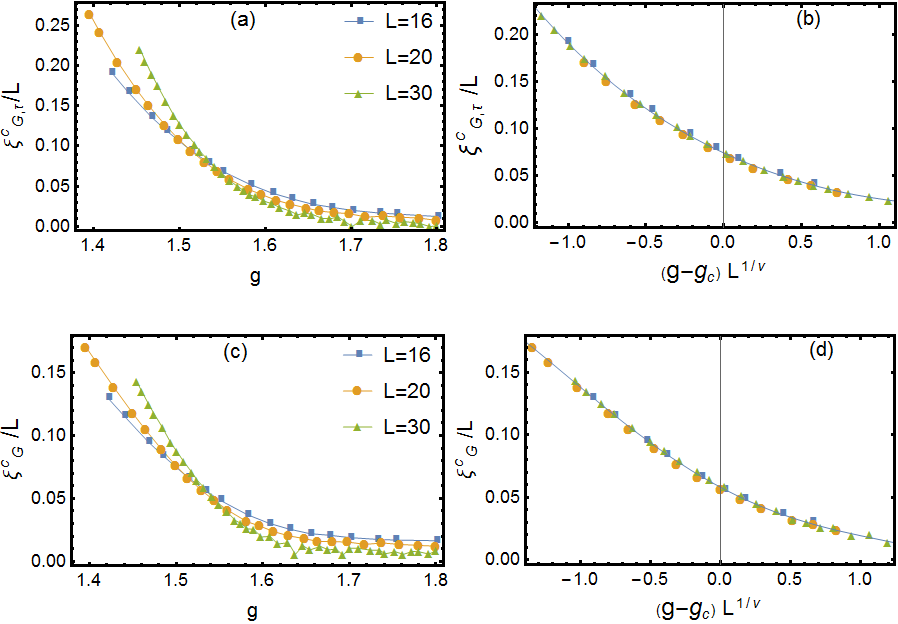}
\caption{(a) Chiral correlation length in the imaginary time direction $\xi^c_{G,\tau} $  of the binary phase-glass model near the CG-I transition at $x=0.2$ for  different system sizes $L$   
and (b) corresponding scaling  plot of the data near the transition with  $g_c=1.54  $,  $\nu = 1.3 $.
(c) Correlation length in the spatial direction $\xi^c_G $ and (d) corresponding scaling  plot   with  $g_c= 1.53 $,  $\nu = 1.3 $. 
$L_\tau=a L^z$, with aspect ratio $a=0.642$ and $z=1.2$}
\label{vcorCGI_pg}
\end{figure}

\begin{figure}
\centering
\includegraphics[width=0.9\columnwidth]{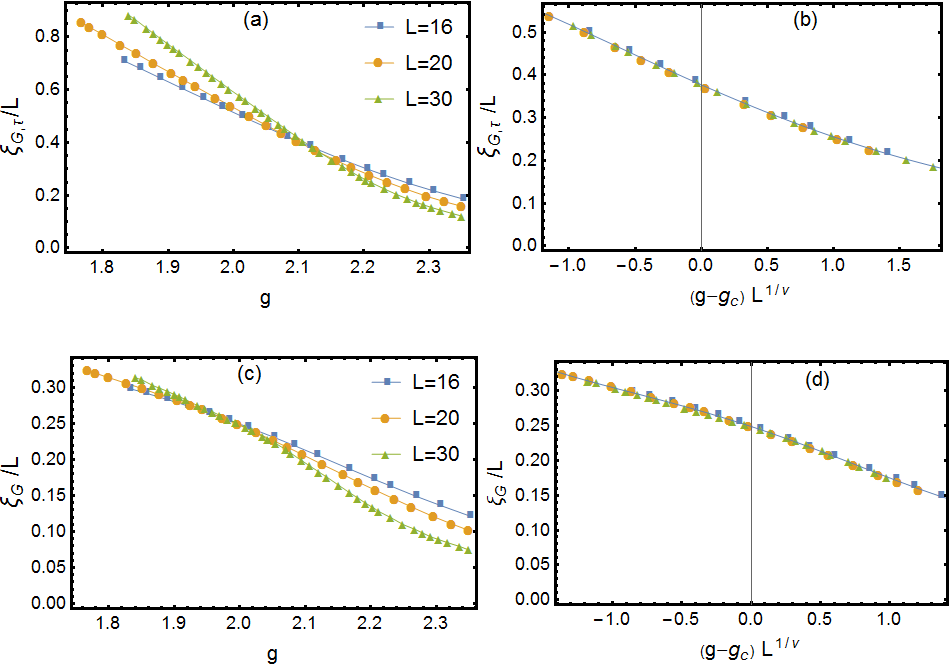}
\caption{(a) Same as  Fig. \ref{corCGI_pg} but for  the Gaussian phase-glass model at $\Delta=1.6$    
and (b) corresponding scaling  plot of the data near the transition with  $g_c=2.123  $,  $\nu = 1.36 $.
(c) Correlation length in the spatial direction $\xi_G $ and (d) corresponding scaling  plot   with  $g_c= 2.0 $,  $\nu = 1.7 $. }
\label{corCGI_gg}
\end{figure}

\begin{figure}
\centering
\includegraphics[width=0.9\columnwidth]{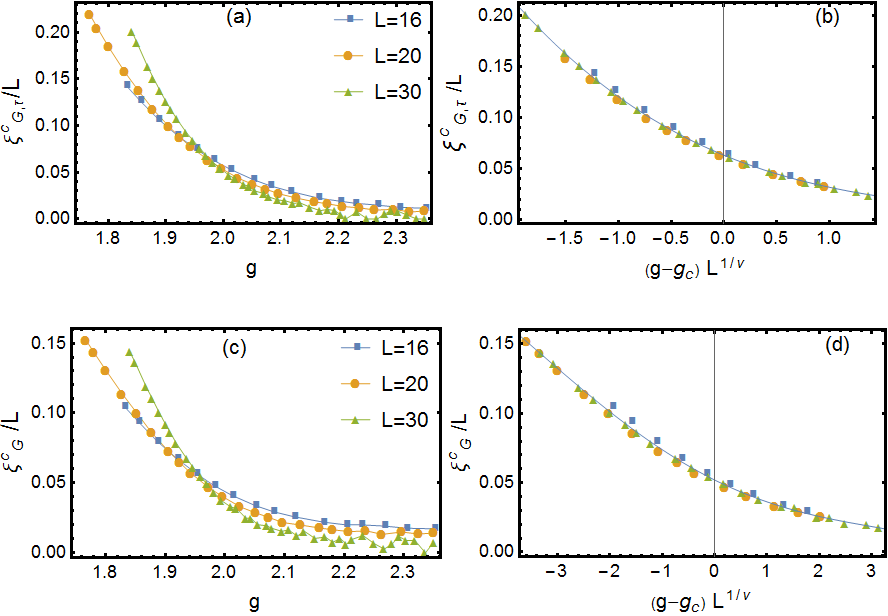}
\caption{(a) Same as  Fig. \ref{vcorCGI_pg} but for  the Gaussian phase-glass model at $\Delta=1.6$    
and (b) corresponding scaling  plot of the data near the transition with  $g_c=1.978  $,  $\nu = 1.3 $.
(c) Correlation length in the spatial direction $\xi^c_G $ and (d) corresponding scaling  plot   with  $g_c= 1.963 $,  $\nu = 1.03 $. }
\label{vcorCGI_gg}
\end{figure}

Figure  \ref{corCGI_pg} shows the behavior of the correlation length  $\xi_{G,\tau} $  and  $\xi_{G}$ in the time and spatial directions, for the binary phase-glass model at $x=0.2$. The dynamic exponent is set to $z=1.2$, the same value estimated for the symmetric models \cite{eg2017random}. We find that this value gives consistent results for the scaling behavior and also agrees with independent estimates of $z$ using large systems, as will be described further ahead. As shown in Fig. \ref{corCGI_pg}a, the curves  for $\xi_{G,\tau}/L $ as a function of $g$ for different system sizes cross approximately at the same point $g_c$, providing evidence of a continuous CG-I transition.  Fig. \ref{corCGI_pg}b,  shows the  scaling plot according to Eq. (\ref{correl}) providing  estimates for  $g_c$ and $\nu$. Figs. \ref{corCGI_pg}c and  \ref{corCGI_pg}d shows the corresponding behavior  for the correlation length in the spatial direction.  The results for the chiral correlation length $\xi^c_{G} $ are shown 
in Figure  \ref{vcorCGI_pg}. The behavior is essentially the same but the data are much noisier. Similar results are obtained for the Gaussian phase-glass model as shown in Fig.  \ref{corCGI_gg} and \ref{vcorCGI_gg}.

\begin{figure}
\centering
\includegraphics[width=\columnwidth]{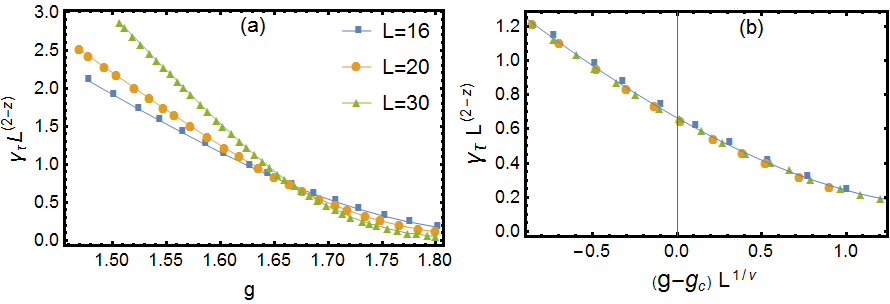}
\caption{ (a) Phase stiffness in the imaginary time direction $\gamma_\tau $ for the binary phase-glass model near the CG-I transition at $x=0.2$ for 
different system sizes $L$ and (b) corresponding 
scaling plot  with  $g_c= 1.675 $  and $\nu =1.2 $.  }
\label{helCGI_pg} 
\end{figure}

\begin{figure}
\centering
\includegraphics[width=\columnwidth]{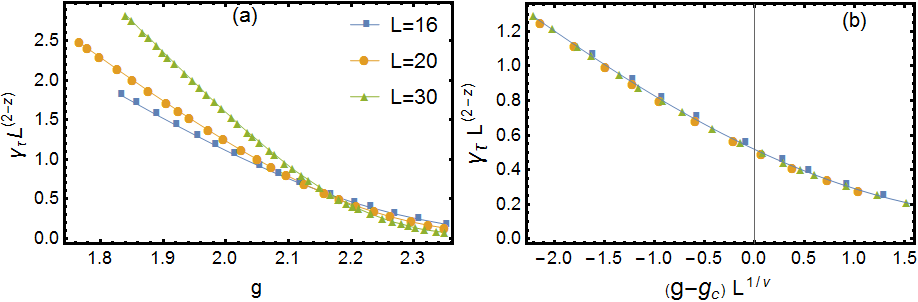}
\caption{ (a) Same as Fig. \ref{helCGI_pg} but for the  Gaussian phase-glass  at $\Delta=1.6$  and (b) corresponding 
scaling plot  with  $g_c= 2.176 $  and $\nu =1.21 $.  }
\label{helCGI_gg} 
\end{figure}

In Fig. \ref{helCGI_pg}, we show the behavior of the phase stiffness in the time direction. Curves for different system sizes  cross  approximately at the same  critical coupling $g_c$ and the corresponding  scaling plot of the data  according to the scaling form of Eq. \ref{rhotau} provide  alternative estimates of  $g_c$  and $\nu$.  The same  behavior  is found for the Gaussian phase-glass model as shown in Figs.  \ref{helCGI_gg}.

It should be noted that there are significant discrepancies in the critical couplings obtained from the scaling behavior of the correlations lengths 
$\xi_G$,  $\xi^c_G$ and phase stiffness $\gamma_\tau$, for both models. They are likely to result from different corrections to finite-size scaling 
and we thus assume they represent estimates of the same  CG-I transition, where the phase-coherence transition is accompanied by a chiral transition and described by a single divergent length scale.
 
\begin{figure}
\centering
\includegraphics[width=\columnwidth]{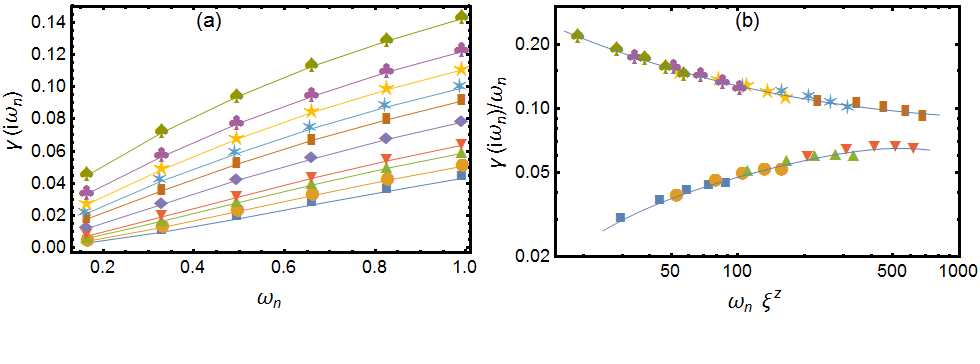}
\caption{ (a) Phase stiffness $\gamma(i\omega)$  for the binary phase-glass model  near the CG-I transition at $x=0.2$, for different couplings $g$ with $L=30$.  From the top bottom: $g=1.50,1.55,1.58,1.61,1.63,1.67,1.71,1.73,1.76$  and $ 1.80 $.
(b) Scaling plot  with $\xi=|g/g_c-1|^{-\nu}$ for   $g_c=1.672$ and  $z \nu=1.75$.  }
\label{helw_pg}
\end{figure}

\begin{figure}
\centering
\includegraphics[width=\columnwidth]{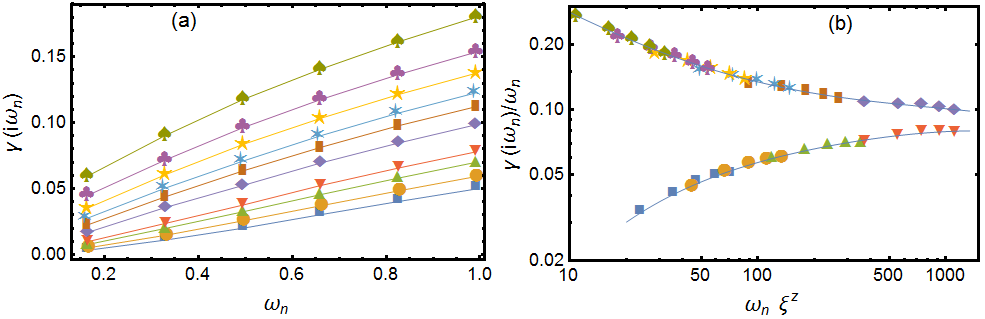}
\caption{ (a) Same as Fig. \ref{helw_pg} but for the Gaussian phase-glass model at $\Delta=1.6$.  From the top bottom: $g=1.84,1.92,1.97,2.02,2.06,2.11,2.19,2.23,2.29$  and $ 2.35 $.
(b) Scaling plot  with $\xi=|g/g_c-1|^{-\nu}$ for   $g_c=2.15$ and  $z \nu=1.8$.  }
\label{helw_gg}
\end{figure}

The scaling behavior described above for the correlation $\xi_G$ and  phase stiffness $\gamma_\tau$ already indicates phase coherence below the CG-I transition and, therefore, the chiral-glass phase is expected to be superconducting. To further investigate the superconducting properties of this phase, we need to look at the phase stiffness in the spatial direction. Unfortunately, the dominant effect of the gauge disorder  leads to negative values for the  phase stiffness \cite{huse90,kostakino99} when obtained directly, as in Eq. (\ref{eqstiff}), depending on the disorder configurations. It turns out, however, that the frequency-dependent phase stiffness $\gamma(i\omega)$ defined in Eq. (\ref{helw}) is well behaved for nonzero frequencies $\omega$.  Its  scaling behavior at small frequencies determines the electrical conductivity from the Kubo formula (Eq. \ref{kubo}).
If  $\gamma(i\omega)$  is finite  when $\omega \rightarrow 0$ in the chiral-glass phase then $\sigma$ diverges and this phase is superconducting.
Near the CG-I transition, it should therefore satisfy the scaling relation \cite{cha3,wallin94}
\begin{equation}
\gamma(i\omega_n)/\omega_n = F_\pm (\omega_n \xi^z),
\end{equation}
in the absence of finite size effects. The + and -  signs correspond to $g  > g_c$ and $g  < g_c$. Indeed, as shown in  Fig. \ref{helw_pg} for the binary and in Fig. \ref{helw_gg} for the Gaussian phase-glass models, the phase stiffness $\gamma(i\omega_n)$ at low frequencies and different  couplings $g$  for a large system where finite-size effects are small  satisfy the above scaling form, providing evidence for a phase transition from a superconducting chiral glass to insulator transition. 

To further verify the finite phase stiffness of the chiral-glass phase and obtain an independent estimate of $z$, we have also studied  the scaling behavior of the phase slippage response ("nonlinear resistivity") $R_x = V_x/J_x$, obtained by driven MC dynamics \cite{eg04,eg2017random} as described in Sec. \ref{MC}. As shown in the Appendix, the scaling analysis is consistent with the above results obtained  from $\xi_G$ and $\gamma(i\omega)$ scaling and therefore provide further support for the finite phase stiffness and a superconducting  chiral-glass phase. 

\begin{figure}
\centering
\includegraphics[width=1.0\columnwidth]{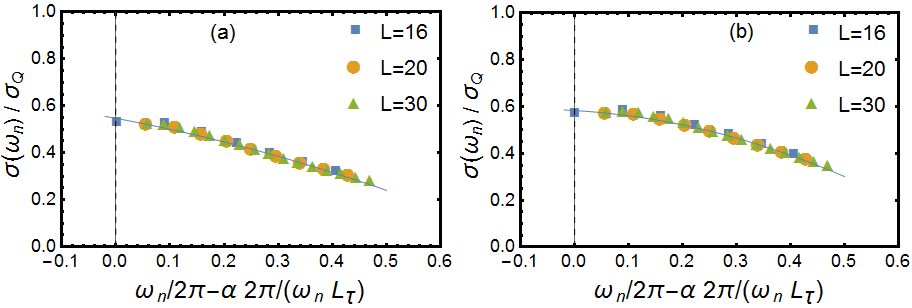}
\caption{  Scaling plot of conductivity $\sigma(iw_n)$ at the critical coupling $g_c$ for (a) the binary phase-glass model at $x=0.2$,
 $g_c=1.672 $ and (b) the Gaussian phase-glass model at $\Delta=1.6$, $g_c=2.15$ with $\alpha=0.06$. The universal conductivity  is given by the intercept with the $x_o=0$ dashed line, leading to the corresponding estimates (a) $\frac{\sigma^*}{\sigma_Q}= 0.55(3)$  and (b) $\frac{\sigma^*}{\sigma_Q}= 0.59(3)$. }
\label{condCGI}
\end{figure}

Since we have found that the chiral-glass phase is superconducting, it is interesting to obtain an estimate of the conductivity $\sigma^*$ at the CG-I transition.  Following the scaling method  of  Cha {et al.} \cite{cha2,cha3} described in Sec. \ref{SI}, $\sigma^*$ can be determined  from a scaling plot according to  Eq. (\ref{cond}).  Fig. \ref{condCGI}a shows this scaling plot  for the  binary phase glass model and Fig. \ref{condCGI}b for the Gaussian phase-glass model, from which we estimate at the CG-I transition  $\sigma^*/\sigma_Q = 0.55(3)$ and $\sigma^*/\sigma_Q = 0.59(3)$, respectively.

\begin{table}[h]
\begin{center}

\begin{tabular}{ c| c c c  }
%\begin{tabular}{ c|c|c|c|c|c}
\hline \hline
 & \multicolumn{3}{c}{binary model}  \\
 \hline \hline
     & S-I  &   CG-I  &    S-CG    \\
   \hline
  $ x_c, g_c $ & $0.05, 2.049(1) $  & $ 0.2, 1.64(4)  $  & $ 0.10(2), 1.6 $    \\
  \hline
  $z$   & $1.10(5)$  & $1.2(1)$ & $1.1(1) $   \\
 \hline
 $\nu$ & $0.85(2)$  &  $1.5(2)$ & $1.5(3)$ \\
 \hline
 $\sigma^*/\sigma_Q$ & $ 0.32(3)$ & $0.55(3)$ &  - \\
 \hline
 \end{tabular}
 
 \begin{tabular}{ c| c c c  }
%\begin{tabular}{ c|c|c|c|c|c}
\hline
 & \multicolumn{3}{c}{Gaussian model}  \\
 \hline \hline
     & S-I  &   CG-I  &    S-CG    \\
   \hline
  $ \Delta_c, g_c $ & $0.5, 2.178(1))$  & $ 1.6, 2.10(5)   $  & $ 1.00(2), 1.68 $    \\
  \hline
  $z$   & $1.10(5)$  & $1.2(1)$ & $1.1(1) $   \\
 \hline
 $\nu$ & $0.85(2)$  &  $1.4(2)$ & $1.4(3)$ \\
 \hline
 $\sigma^*/\sigma_Q$ & $ 0.31(3)$ & $0.59(3)$ &  - \\
 \hline
 \end{tabular}

 \caption{\label{tab:disor} Critical couplings $(x_c,g_c)$ or $(\Delta_c,g_c)$,  exponents $z, \nu $ and 
 conductivity $\sigma^*$ at the transition of the S-I, CG-I and S-CG transitions in the phase diagrams of Fig. \ref{phd},
 for the binary and Gaussian phase-glass models. At the S-CG transition $\sigma$ remains infinity. }
\end{center}
\end{table}

\section{Summary and Conclusions}

We have studied the superconductor to insulator transition in two-dimensional phase-glass
(or chiral-glass) models with varying degree of disorder by path-integral MC simulations and finite-size scaling. 
Two different models were considered, with binary and  Gaussian distribution of quenched disorder, 
having nonzero mean. Both models display the same topology of the phase diagram (Fig. \ref{phd}) 
with critical exponents that agree within the estimated errorbars (Table I). 
In addition to the usual superconducting and insulating phases, a chiral-glass phase  occurs for sufficiently large disorder. 
The chiral-glass to insulator transition is in a  different universality class, with critical exponents and universal conductivity at the transition significantly different from those of the superconductor-insulator transition for small disorder. 

The transition from superconductor to insulator can take place via the intermediate chiral-glass phase,  depending on the parameters of the models. We find, however, that the chiral-glass phase has a finite phase stiffness, being still a superconductor,  instead of the Bose metal, which has been  suggested by the mean-field theory approach \cite{dalidovich2002phase,phillips2003absence,phillips03}. This indicates that the 2D phase-glass model does not provide a theoretical framework for the Bose metal state in the zero temperature limit observed experimentally  \cite{jaeger1989,yazdani1995,chervenak2000,christiansen2002evidence,tsen2016}, at least in its simplest form with onsite charging energy, short-range Josephson interactions, and an absence of magnetic screening effects. 

Our results are also relevant for superconducting thin films nanostructured with a periodic pattern of nanoholes  and doped with magnetic impurities \cite{ zhang2019quasiparticle}. Assuming that these magnetic impurities  introduce $\pi$ junctions \cite{bula77}  distributed randomly, the transition to the insulating phase for sufficient large impurity doping should be in the universality class of the chiral-glass to insulator transition studied here. 
The disappearance of magnetic frustration effects for the activation energy in the insulating phase at large doping \cite{ zhang2019quasiparticle}, is in fact consistent with such a chiral-glass  regime. 
%\acknowledgements
\begin{acknowledgments}
The author thanks  J. M. Kosterlitz  and J. M. Valles, Jr. for helpful discussions.
This work was supported by Funda\c c\~ao de Amparo \`a Pesquisa do Estado de S\~ao Paulo - FAPESP (Grant No. 18/19586-9),  CNPq (Conselho Nacional de Desenvolvimento Cient\'ifico e Tecnol\'ogico)  and computer facilities from CENAPAD-SP.
\end{acknowledgments}

\section{Appendix}

\begin{figure}
\centering
\includegraphics[width=\columnwidth]{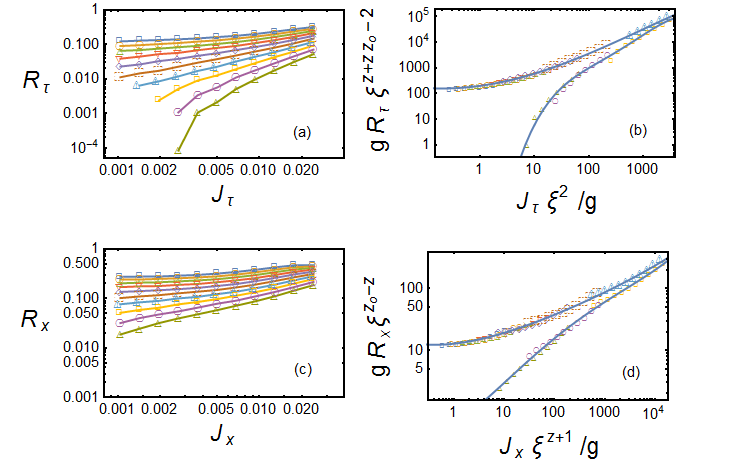}
\caption{ Scaling behavior of the  phase slippage response for the binary phase-glass model  near the CG-I transition at $x=0.2$ for $L=60$ in (a) the imaginary-time direction $R_\tau$ and (c) spatial direction $R_x$ near the CG-I transition. From the top down, the couplings are $g=1.80, 1.78, 1.76, 1.74, 1.72, 1.70, 1.68, 1.66, 1.64 $  and $ 1.62 $.
(b) and (d) Scaling plots corresponding to (a) and (c), respectively, for data near the transition with $\xi=|g/g_c-1|^{-\nu}$ using the same 
parameters $g_c=1.67$, $ z_o=2.3$, $z=1.25$, and $\nu = 1.2$. }
\label{Rx_pg}
\end{figure}

Here  we consider the scaling behavior of the phase slippage response ("nonlinear resistivity") $R_x = V_x/J_x$  near the CG-I transition, obtained by driven MC dynamics \cite{eg04,eg2017random} as described in Sec. \ref{MC}. 
%Similarly,
%we define $R_\tau$ as the phase slippage response to the applied perturbation $J_\tau$ in the layered (imaginary-time) direction.
%The corresponding linear "conductivity" $\sigma = 1/R_x|_{J_x \rightarrow 0}$, is a measure of the phase stiffness of the system $\gamma_x$ througth the relation $\sigma(w)=\gamma_x(w)/i w $, where $w$ is the frequency of the MC dynamics. 
If the phase stiffness is finite in the chiral-glass phase, we then expects that  $R_x$ should approach a nonzero value when $J_x \rightarrow 0$ above the transition, $g > g_c$,  while below the transition it should  approach zero. From the nonlinear scaling behavior near the transition \cite{eg2017random,girvin} of a sufficiently large system, one can also extract the critical coupling $g_c$, and the critical exponents $\nu$ and $z$. 
Figs. \ref{Rx_pg} shows the behavior of the nonlinear phase slippage response $R_x$ and $R_\tau$ for the binary  phase-glass model as a function of the applied perturbation $J_x$ and $J_\tau$, respectively. The behavior for different values of $g$ is consistent with a phase-coherence transition at an apparent critical coupling in the range $g_c \sim 1.68 - 1.66$. For $g > g_c$, both
$R_x$ and $R_\tau$ tend to a finite value while for $g < g_c$, they
extrapolate to low values. Assuming the transition is continuous,
the nonlinear response behavior sufficiently close to the transition should satisfy a scaling form in terms of $J_x$, $J_\tau$ and $g$.
The critical coupling $g_c$ and critical exponents $\nu$ and $z$ can then be obtained from the
best data collapse satisfying the scaling behavior close to the transition. $R_x$  and $R_\tau$ should satisfy the scaling forms \cite{girvin},
\begin{eqnarray}
g R_x \xi^{z_0-z} & = & F_\pm (J_x \xi^{z+1}/g),  \cr
g R_\tau \xi^{z + z_0 z - 2} & = & H_\pm (J_\tau \xi^2/g),
\label{Rxy}
\end{eqnarray}

where $z_o$ is an additional critical exponent describing the MC relaxation times, $t^r_{mc, x}\sim \xi^{z_o}$ and  $t^r_{mc,\tau}\sim \xi_\tau^{z_o}$, in the spatial and imaginary-time directions, respectively,  and  $\xi=|g/g_c-1|^{-\nu}$.
The + and -  signs correspond to $g  > g_c$ and $g  < g_c$.  
The joint scaling plots according to Eqs. (\ref{Rxy}) are shown in  Figs. \ref{Rx_pg}b and  \ref{Rx_pg}d, obtained by adjusting the unknown parameters. This scaling analysis gives the estimates $g_c=1.670(5) $, $\nu=1.10(5)$
and  $z=1.25(5) $, which are consistent with the results from the correlation-length scaling described in Sec. \ref{CGI} and therefore provide further support for the finite phase stiffness of the chiral glass phase. 

%\bibliography{bibpg}

 \newcommand{\noop}[1]{}

\end{document}